\documentclass[3p,times,twocolumn]{elsarticle}

\makeatletter
\def\@author#1{\g@addto@macro\elsauthors{\normalsize%
    \def\baselinestretch{1}%
    \upshape\authorsep#1\unskip\textsuperscript{%
      \ifx\@fnmark\@empty\else\unskip\sep\@fnmark\let\sep=,\fi
      \ifx\@corref\@empty\else\unskip\sep\@corref\let\sep=,\fi
      }%
    \def\authorsep{\unskip,\space}%
    \global\let\@fnmark\@empty
    \global\let\@corref\@empty  
    \global\let\sep\@empty}%
    \@eadauthor={#1}
}
\makeatother

\makeatletter
\def\ps@pprintTitle{%
 \let\@oddhead\@empty
 \let\@evenhead\@empty
 \def\@oddfoot{}%
 \let\@evenfoot\@oddfoot}
\makeatother

\usepackage{amssymb}

\usepackage[figuresright]{rotating}

\begin{document}

\begin{frontmatter}


\title{Backgrounds and sensitivity of the NEXT double beta decay experiment}

\author{M. Nebot-Guinot\corref{cor1}\fnref{ific}}
\cortext[cor1]{Corresponding author e-mail: miquel.nebot@ific.uv.es}

\author{P. Ferrario\fnref{ific}}
\author{J. Mart\'in-Albo\fnref{ific}}
\author{J. Mu\~noz Vidal\fnref{ific}}
\author{J.J. G\'omez-Cadenas \fnref{ific}}
\author{ \\ on behalf of the NEXT Collaboration} 
\address[ific]{Instituto de F\'isica Corpuscular (IFIC), CSIC \& Universitat de Val\`encia \\Calle Catedr\'atico Jos\'e Beltr\'an, 2, 46980 Paterna, Valencia, Spain\\\fnref{label3}}

\begin{abstract}
NEXT (Neutrino Experiment with a Xenon TPC) is a neutrinoless double-beta (\(\beta\beta0\nu\)) decay experiment that will operate at the Canfranc Underground Laboratory (LSC). It is an electroluminescent high-pressure gaseous xenon  Time Projection Chamber (TPC) with separate read-out planes for calorimetry and tracking. Energy resolution and background suppression are the two key features of any neutrinoless double beta decay experiment. NEXT has both good energy resolution (\(<1\%\) FWHM) at the Q value of $^{136} $Xe and an extra handle for background identification provided by track reconstruction.
	With the background model of NEXT, based on the detector simulation and the evaluation of the detector radiopurity, we can determine the sensitivity to a measurement of the $\beta\beta2\nu$ mode in NEW and to a \(\beta\beta0\nu\) search in NEXT100. In this way we can predict the background rate of \(5\times10^{-4}\) counts/(keV kg yr), and a sensitivity to the Majorana neutrino mass down to 100 meV after a 5-years run of NEXT100.
\end{abstract}


\end{frontmatter}


\section{Neutrinoless double beta decay}
\label{intro}
Neutrinoless double beta decay ($\beta\beta0\nu$) is a postulated nuclear transition in which two neutrons undergo $\beta$ decay simultaneously without the emission of neutrinos. 

 Evidence of this process would establish that massive neutrinos are Majorana particles, provide a hint of a new physics scale beyond the Standard Model and prove the violation of total lepton number, a key element to explain the observed asymmetry between matter and antimatter in the universe. 
In addition, the measurement of the $\beta\beta0\nu$-decay rate would provide information on the absolute scale of neutrino masses \cite{GomezCadenas:2011it}, as shown in Eq.\ref{eq1}:
\begin{equation} 
\left ( T^{0\nu}_{1/2} \right )^{-1 }\propto m^{2}_{\beta\beta}
\label{eq1}
\end{equation}

\section{Double beta decay experiments}
\label{experiments}
Double beta decay detectors  measure the sum of the kinetic energies from the two released electrons, $Q\beta\beta$. Considering the finite energy resolution ($\Delta E$) of any detector, other processes occurring in the detector, as the tail of the $\beta\beta2\nu$ mode, can fall in the region of energies around $Q_{bb}$ becoming background. As in other rare event detectors, backgrounds of cosmogenic origin and natural radioactivity from materials are a problem, and thus underground operation and selection of radiopure materials   is essential. In this sense additional experimental features are desired to improve the sensitivity of the detector, such as extra background (B) rejection, better detector efficiency ($\epsilon $) or larger exposure (${M\cdot t} $) \cite{GomezCadenas:2011it}. This relation can be summarized as follows :
\begin{equation} 
T_{1/2} \propto a\cdot \epsilon \sqrt{\frac{M\cdot t}{\Delta E \cdot B}} 
\end{equation}

\begin{table*}[t]
\centering
\resizebox{\textwidth}{!} {
\begin{tabular}{l l l l c c c c r} 
\hline
Component & Material &  Unit & Quantity (in NEW) & $^{208}$Tl & $^{214}$Bi & $^{40}$K & $^{60}$Co  & Technique \\ 							
\hline \hline
Dice Boards	&		&	Bq/Unit 	& 28 Units 	& 	4.00E-05		&	3.00E-05		&	1.21E-02	&	1.00E-05 	& NEXT\\
Field Cage	&	HDP	&	Bq/kg	&18.626 kg	&	7.56E-06		&	6.20E-05		&	$<$3.40E-03&	$<$1.40E-04 & ICPMS/NEXT\\
			& Resistors& 	Bq/Unit	&106 Units	&	2.52E-06		&	1.64E-05		&	1.90E-05	&	1.10E-06	& DarkSide\\
ICS 			& CuA1	&	Bq/kg	&651.031 kg	&	1.44E-06		&	1.20E-05		&	3.70E-04	&	4.10E-05	& GDMS /NEXT\\	
PMT Body		&		&	Bq/Unit	&12 Units		&	1.44E-04		&	5.00E-04		&	1.39E-02	&	4.40E-03	& XENON/NEXT \\
Vessel		&316Ti SS & 	Bq/kg	&606,005 kg	&	1.48E-04		&	4.60E-04		&	1.20E-04	&	4.40E-03 	& GDMS\\
Carrier Plate	&CuA1	&	Bq/kg	&239.607 kg 	&	1.44E-06		&	1.20E-05		&	3.70E-04	&	4.10E-05 	&GDMS/NEXT\\	
Support Plate 	&CuA1	&	Bq/kg	&272.614 kg	&	1.44E-06		&	1.20E-05		&	3.70E-04	&	4.10E-05	&GDMS/NEXT\\ Enclosure Body&CuA1	&	Bq/Kg	&79.941 kg	&	1.44E-06		&	1.20E-05		&	3.70E-04 	&	4.10E-05	&GDMS/NEXT\\
Enclosure Window&Sapphire&Bq/Kg	&1.654 kg		&	$<$1.98E-03	&	$<$2.70E-03	&	$<$1.80E-02&	$<$7.00E-04	&NEXT\\
Shielding Lead	 &Lead	&	Bq/Kg	& 15614.7 kg	&	3.39E-05		&	3.47E-04		&	1.24E-04	&	9.00E-05	&GDMS\\
Pedestal		&316TiSS&	Bq/Kg	&360 kg		&	1.48E-04		&	4.60E-04		&	1.20E-04	&	4.40E-03	&GDMS\\
Cu castle		&CuA1	&	Bq/Kg	&4056.568 kg	&	1.44E-06		&	1.20E-05		&	3.70E-04	&	4.10E-05 	&GDMS/NEXT\\			
\hline
\end{tabular}
}
\caption{Radioactive budget of the detector components of NEXT. }
\label{RBudget}
\end{table*}

\section{NEXT-100}
\label{next100}

The NEXT-100 detector will search for the neutrinoless double beta decay of $^{136}$Xe at the  Laboratorio Subterraneo de Canfranc. It uses a time projection chamber filled with 100 kg of enriched xenon gas at 15 bar pressure, with separated detection functions for calorimetry and tracking \cite{Alvarez:2012haa}.
The gaseous xenon provides scintillation and ionization as primary signals. These are used to establish the start-of-event time ($t_0$) and for calorimetry/tracking respectively. In order to achieve optimal energy resolution, the ionization signal is amplified in NEXT using the electroluminescence (EL) of xenon \cite{Nygren:2009zz}.

\emph{Calorimetry :} The energy plane is made of 60 photomultiplier tubes (Hamamatsu R11410-10 PMTs), located behind the TPC. 
These PMTs will be sealed into individual pressure resistant, vacuum tight, copper enclosures coupled to sapphire windows in order to withstand the  high pressure of the gas \cite{Alvarez:2012haa}.

\emph{Tracking : }The tracking will be provided by a dense array of $1~\mathrm{mm}^2$ silicon photomultipliers (Hamamatsu S10362-11-050P SiPMs), measuring the forward-going secondary scintillation. It is located behind the anode, very close to the EL region, and is used for event topological reconstruction \cite{Alvarez:2012haa}.

In the first phase of underground operation NEXT will use a 1:2 scale prototype  that we call NEXT-White (NEW).
We refer the reader to the dedicated paper in these proceedings \cite{Monrabal}.

\section{Background model}
\label{backroundmodel}
A detailed simulation of the detector performance has been implemented in NEXUS, the Geant4-based simulation program of the NEXT experiment. This let us evaluate all the main backgrounds that can mask the $\beta\beta0\nu$ signal, coming from the activities of the materials to be used in the construction (Table \ref{RBudget}). The NEXT Collaboration is carrying out a thorough campaign of material screening and selection with the assistance of the LSC Radiopurity Service \cite{Alvarez:2012as}. 

\begin{figure}[h!]
\centering
\includegraphics[width=0.4\textwidth]{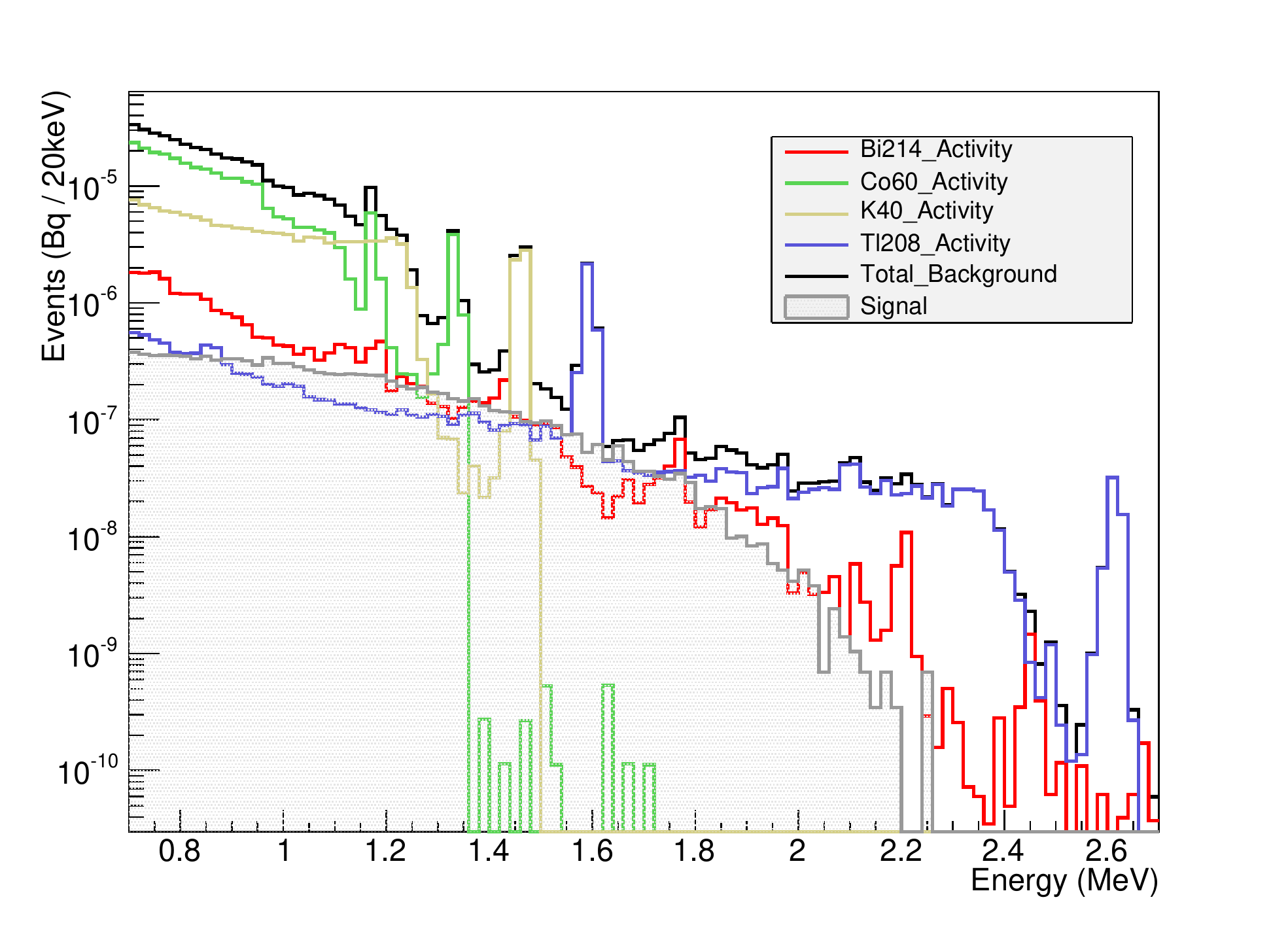}
\caption{Background spectra expected in NEW differentiated by source and compared to the $\beta\beta2\nu$ signal rate}
\label{NEW}
\end{figure}

Figure \ref{NEW} shows the expected background and $\beta\beta2\nu$ signal in the NEW prototype. With the operation and first results of NEW the accuracy of this  model will be validated and improved if needed. 
Using this model the background rate expected in NEXT-100 is estimated to be as good as $5\times10^{-4}$ counts/(keV kg yr) \cite{GomezCadenas:2012jv}.

\section{Analysis and Sensitivity}
\label{sens}
\begin{figure} [h!]
\centering
\includegraphics[width=0.35\textwidth]{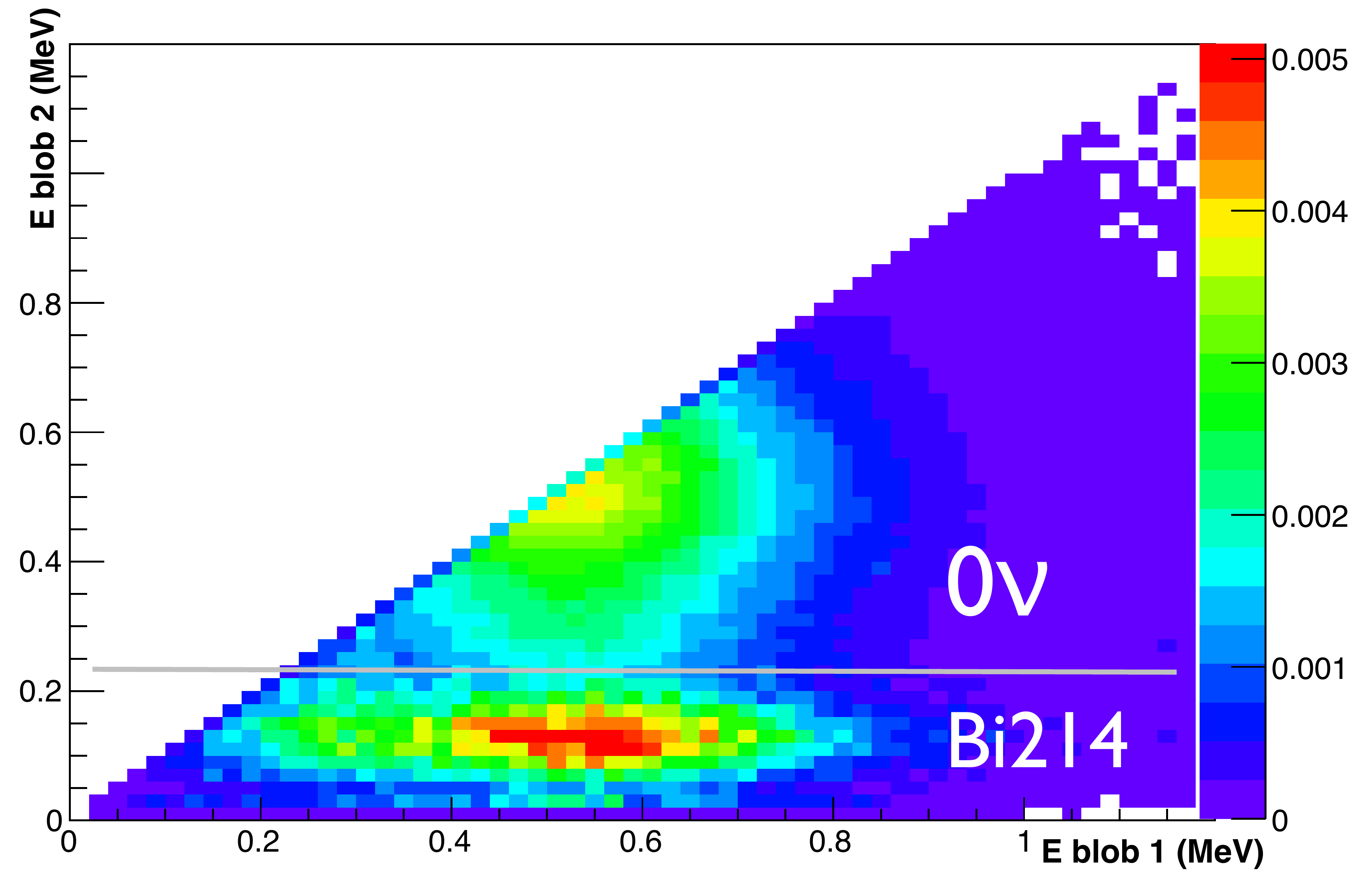}
\caption{Energy of one \emph{blob} vs energy of the other \emph{blob} for signal ($0\nu$) and background (Bi214) events.}
\label{blobs}
\end{figure}

The output of the simulation are \emph{hits} of energy deposited in the active volume of the detector. A standard analysis of those simulated events produced by the modeling has been developed. This analysis has been implemented in the same way as will be done for the real data based on the features that NEXT offers taking into account the detection and reconstruction effects. Here is a brief description of the selection algorithms applied to the events made by the analysis :
\begin{itemize}
\item A first raw cut on event's energy is made to reduce the amount of data produced by the simulation. Only those events with energy greater than 0.6 MeV are written in the preprocessing. 
\item $\beta\beta$ events are produced in the active volume, while background events come from the materials enclosing it.  Therefore, only those events that are fully contained in an inner fiducial volume, far enough from the walls, are selected.
\item Detection and reconstruction effects are taken into account by smearing the event true energy according to the expected resolution of the detector. 
\item A more precise cut on the energy of the events is made selecting only those that enter in our region of interest  (ROI) window. That window is defined before in order to maximize the sensitivity and change among the $2\nu$ or the $0\nu$ analysis.
\item One of the features of NEXT is the ability to reconstruct event topology. The simulated events are reconstructed with the \emph{voxelization} algorithm (looks for a finite space volume with an energy deposition different from zero) to $1~\mathrm{cm}^{3}$  3D hits. Afterward those hits are interconnected forming the track produced by the particle in the detector. Those tracks are analyzed comparing their associated end \emph{blob} (high energy deposition in a small region). The event tracks with one \emph{blob} (background-like) are discarded and those with two \emph{blobs} (signal-like) at the ends are selected.
In Figure \ref{blobs} is clearly shown the different fingerprint made by background and signal like events, that allow us to use this criteria to enhance our sensitivity.
\end{itemize}

\begin{table}
\centering
\resizebox{0.5\textwidth}{!} {
\begin{tabular}{l *{4}{c}}
\hline
Selection cut & \multicolumn{4}{c}{Fraction of events} \\ 
		& $\beta\beta0\nu$ & $\beta\beta2\nu$ &   $^{214}Bi$ & $^{208}Tl$\\	
\hline
$E \in (2.3,2.6)~{\rm MeV}$ & 0.776 & $3.31\times10^{-6}$  & $1.52\times10^{-4}$ & $8.02\times10^{-3}$ \\
Fiducial                    & 0.678 & $2.95\times10^{-6}$  & $1.13\times10^{-4}$ & $4.77\times10^{-3}$ \\
Single track                & 0.508 & $2.27\times10^{-6}$  & $1.36\times10^{-5}$ & $8.44\times10^{-4}$ \\
d$E/$d$x$                   & 0.381 & $1.70\times10^{-6}$  & $1.36\times10^{-6}$ & $8.10\times10^{-5}$ \\
ROI & 0.319 & $3.24\times10^{-12}$ & $1.23\times10^{-7}$ & $3.23\times10^{-7}$ \\
\hline
\end{tabular}
}
\caption{Acceptance of the selection cuts for signal ($\beta\beta0\nu$) and backgrounds.}
\label{cuts}
\end{table}
 
 With the aforementioned rejection factors  (Table \ref{cuts}) and the background model, the half-life sensitivity of  NEXT-100 is estimated to be $1.1\times10^{26} $ years, corresponding to an effective neutrino mass of  $\sim100 $ meV, after 5 years running (500 kg$\cdot$y of exposure)
\cite{GomezCadenas:2012jv}.
 
\begin{figure}[h!]
\centering
\includegraphics[width=0.4\textwidth]{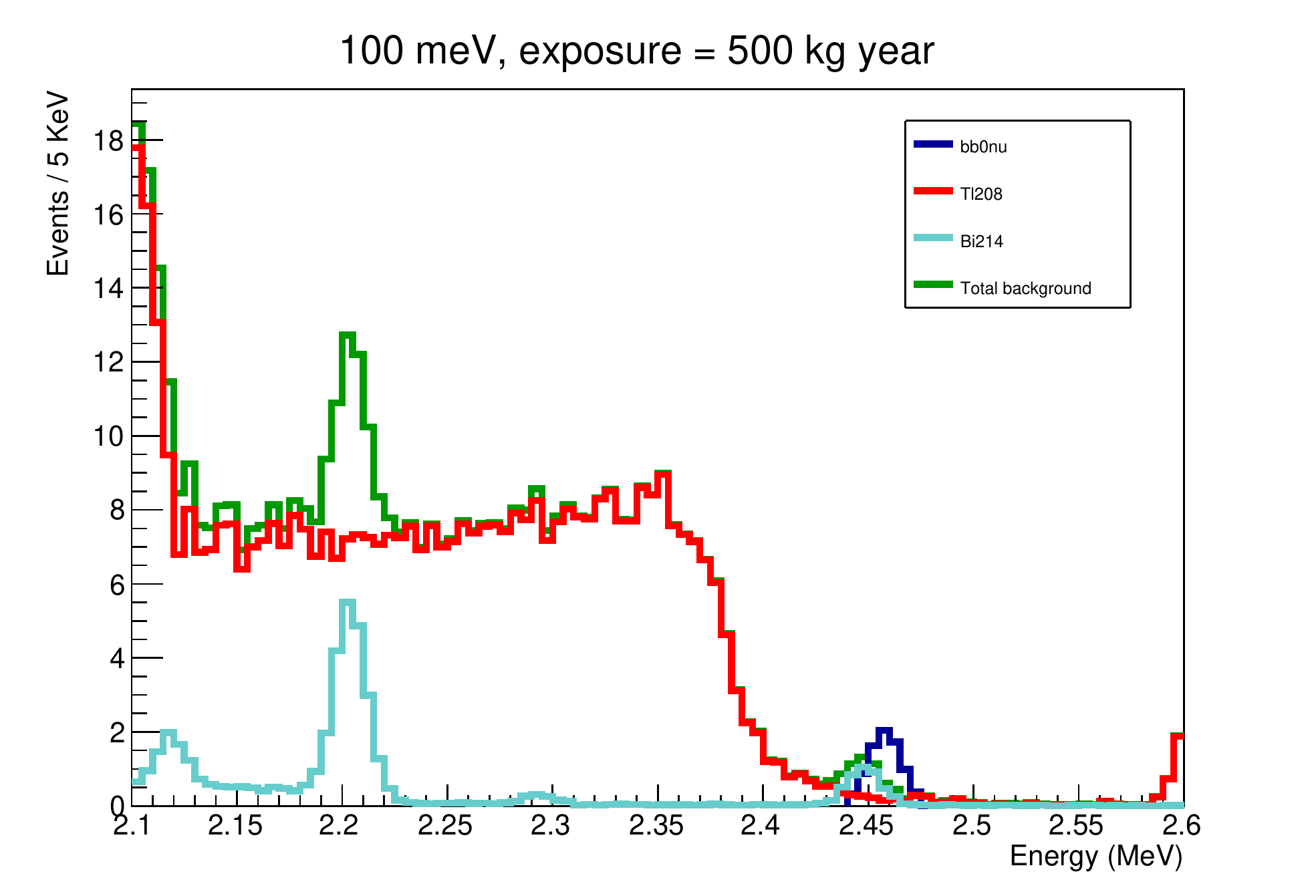}
\caption{NEXT-100 simulated $\beta\beta0\nu$ signal corresponding to a mass of 100 meV and backgrounds entering into the region of interest.} 
\label{NEXT100}
\end{figure}

\section*{Acknowledgement}
The authors would like to acknowledge the support of the Ministerio de Econom'a y Competitividad of Spain under grants CONSOLIDER-Ingenio 2010 CSD2008-0037 (CUP), FPA2009-13697-C04-04 and FIS2012-37947-C04. Also the support of the European Research Council advanced grant 339787 - NEXT.



\nocite{*}
\bibliographystyle{elsarticle-num}
\bibliography{references}







\end{document}